\documentclass[a4paper,twoside,10pt]{article}
\usepackage{graphicx,publaob}

\def\papertype{Progress Report}
\begin{document}

\title{MULTI-BAND INTRA-NIGHT VARIABILITY OF THE BLAZAR CTA\,102 DURING ITS 2016 DECEMBER OUTBURST}

\authors{B. M. MIHOV \lowercase{and} L. S. SLAVCHEVA-MIHOVA}


\address{Institute of Astronomy and NAO, Bulgarian Academy of Sciences, 72 Tsarigradsko Chaussee Blvd. 1784 Sofia, Bulgaria}
\Email{bmihov}{astro.bas}{bg}





\markboth{\runningfont MULTI-BAND INTRA-NIGHT VARIABILITY OF THE BLAZAR CTA\,102}
{\runningfont B. M. MIHOV \lowercase{and} L. S. SLAVCHEVA-MIHOVA}

\vskip-1mm

\abstract{During 2016 December the blazar CTA\,102 underwent an unprecedented outburst thus becoming the brightest blazar observed up to date. We present some results from the intra-night monitoring of the blazar in the $BVRI$ bands in three consecutive nights during the outburst.}

\vskip-.5cm

\section{INTRODUCTION}

\noindent The blazar class of active galactic nuclei involves BL Lacertae objects and flat-spectrum radio-quasars. Violent variability across the electromagnetic spectrum is among their main characteristics. In particular, variability time scales could provide valuable information about the emitting region parameters.

\indent The flat-spectrum radio-quasar CTA\,102 ($z = 1.032 \pm 0.003$) underwent an unprecedented outburst in 2016 December reaching an $R$ band magnitude of 10.82 $\pm$ 0.04 thus becoming the brightest blazar up to date. The outburst was monitored by the GASP--WEBT collaboration, and the results were published in Raiteri et al. (2017).

\indent We perform temporal analysis of the multi-band intra-night variations of CTA\,102 during the outburst with the aim to derive some physical parameters of the emitting regions.

\vskip-3mm

\section{OBSERVATIONS AND PHOTOMETRY}

\noindent We monitored CTA\,102 in the $BVRI$ bands on 2016 December 3$^{\rm rd}$ to 5$^{\rm th}$ for about 4 hours per night using the 50/70\,cm Schmidt telescope of the Rozhen NAO, Bulgaria, and the FLI PL18603 CCD camera (Kostov 2010).

\begin{figure}[ht!]
\centering
\includegraphics[width=0.9\textwidth,keepaspectratio=true]{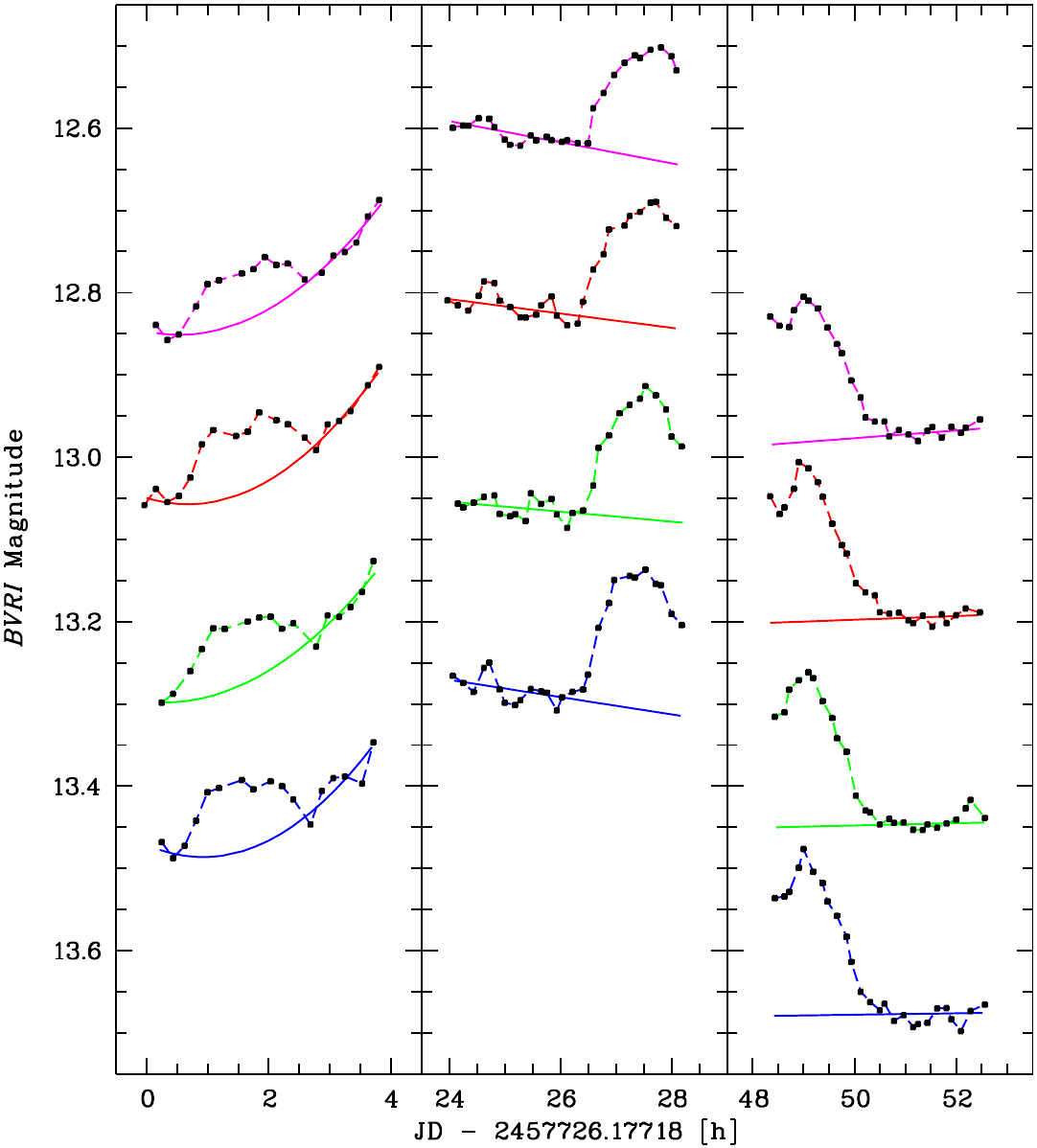}
\vspace{-3mm}
\caption{Light curves of CTA\,102 in $BVRI$ bands (ordered from bottom to top, respectively) for the nights of monitoring. The $VRI$ band LCs are shifted down by 0.40, 0.59, and 0.94\,mag, respectively, for a better display. The error bars are not shown, their sizes are comparable with the symbol sizes. The solid lines depict the fits to the smooth variability component for each band and night.}
\label{fig:lc}
\vspace{-1mm}
\end{figure}

\indent The aperture photometry of the blazar, a control star (a field star of compatible brightness), and a couple of GASP--WEBT suggested reference stars was performed by means of DAOPHOT. The aperture radii were set to 2 or 3 times the FWHM. The light curves (LCs) for the nights of monitoring are shown in Figure\,\ref{fig:lc}. Some characteristics of the obtained intra-night data sets are listed in Table\,\ref{tab:chars}.

\begin{table}
\caption{{Characteristics of the intra-night data sets.}}
\vskip2mm
\centerline{
\resizebox{\linewidth}{!}{
    \begin{tabular}{@{}cccccccc@{}}
			\hline\noalign{\smallskip}
			Date & Band & Monitoring & Time     & Wtd Mean  & Wtd RMS  & Variability & Fractional  \\
            in   &      & Duration   & Sampling & Magnitude & Scatter  & Amplitude   & Variability \\
            2016 &      & [\,h\,]    & [\,min\,]& [\,mag\,] & [\,mag\,]& [\,\%\,]    &             \\
			\noalign{\smallskip}\hline\noalign{\smallskip}
			Dec 3 & $B$ & 3.47 & 12.1 & 13.417 & 0.036 & 14.000 & 0.033 \\
			       & $V$ & 3.47 & 12.1 & 12.815 & 0.038 & 17.186 & 0.030 \\
			       & $R$ & 3.79 & 12.1 & 12.396 & 0.044 & 16.780 & 0.032 \\
			       & $I$ & 3.68 & 12.1 & 11.840 & 0.041 & 17.000 & 0.026 \\
            \noalign{\smallskip}
			  Dec 4 & $B$ & 4.08 & 9.4  & 13.242 & 0.060 & 17.027 & 0.065 \\
			       & $V$ & 4.07 & 9.4  & 12.620 & 0.056 & 17.188 & 0.059 \\
			       & $R$ & 4.07 & 9.4  & 12.187 & 0.052 & 14.972 & 0.057 \\
			       & $I$ & 4.07 & 9.4  & 11.637 & 0.044 & 11.892 & 0.052 \\
            \noalign{\smallskip}
		    Dec 5 & $B$ & 4.09 & 9.4  & 13.613 & 0.075 & 22.096 & 0.071 \\
			       & $V$ & 4.10 & 9.4  & 12.980 & 0.072 & 19.132 & 0.068 \\
			       & $R$ & 4.09 & 9.4  & 12.538 & 0.071 & 19.976 & 0.068 \\
			       & $I$ & 4.09 & 9.4  & 11.965 & 0.066 & 17.490 & 0.065 \\
			\noalign{\smallskip}\hline\noalign{\smallskip}
    \multicolumn{8}{c}{{\em Note.} The fractional variability is computed for the deboosted LCs (see Section\,3).}
	\end{tabular}}}
    \label{tab:chars}
\end{table}

\begin{figure}[ht!]
\centering
\includegraphics[width=0.9\textwidth,keepaspectratio=true]{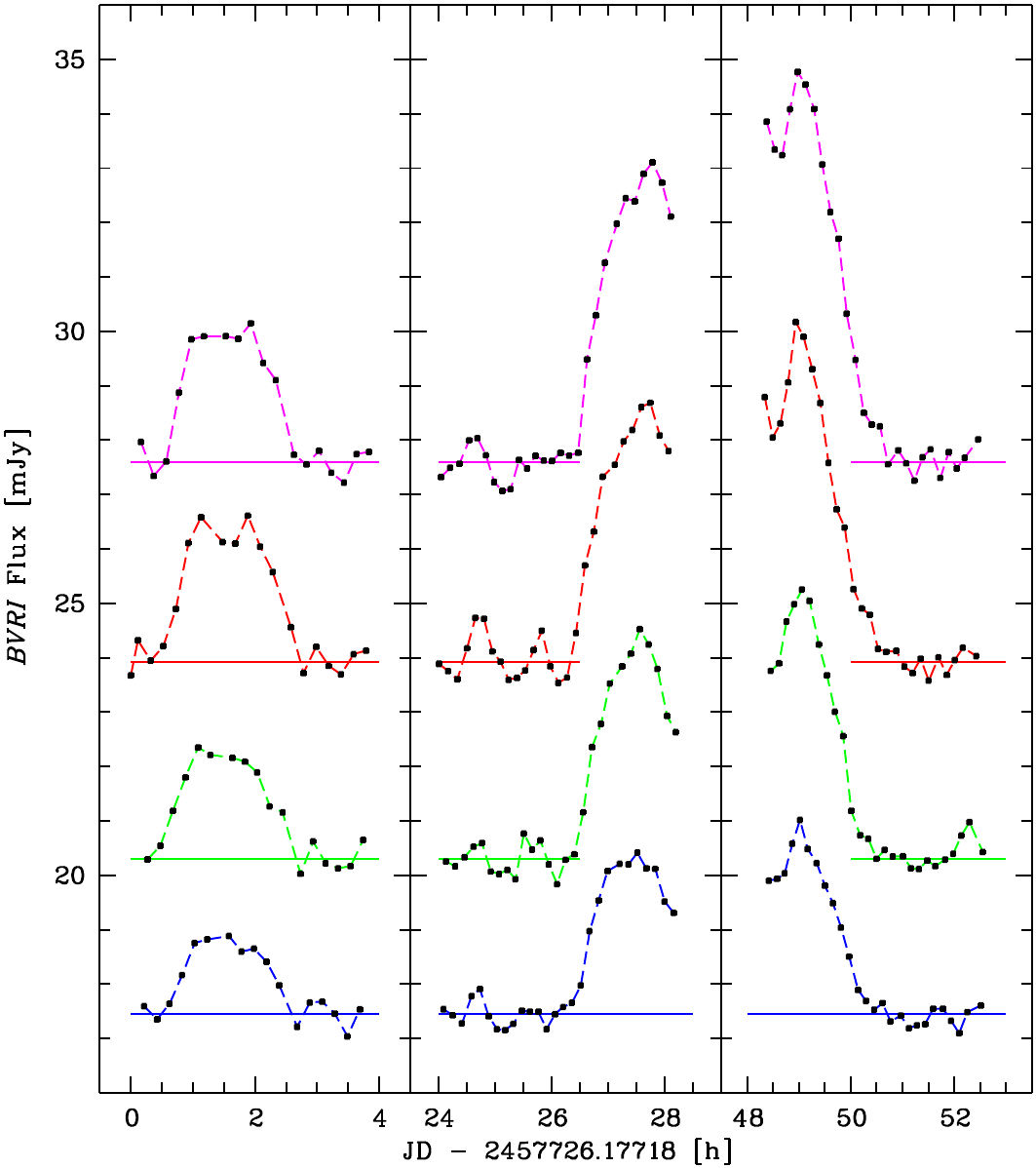}
\vspace{-3mm}
\caption{Deboosted $BVRI$ band LCs (from bottom to top, respectively). Horizontal lines mark the baseline flux level. The $VRI$ band LCs are shifted down by 6, 8, and 13\,mJy, respectively, for a better display.}
\label{fig:lc1}
\vspace{-1mm}
\end{figure}

\begin{figure}[t]
\centering
\includegraphics[width=0.9\textwidth,keepaspectratio=true]{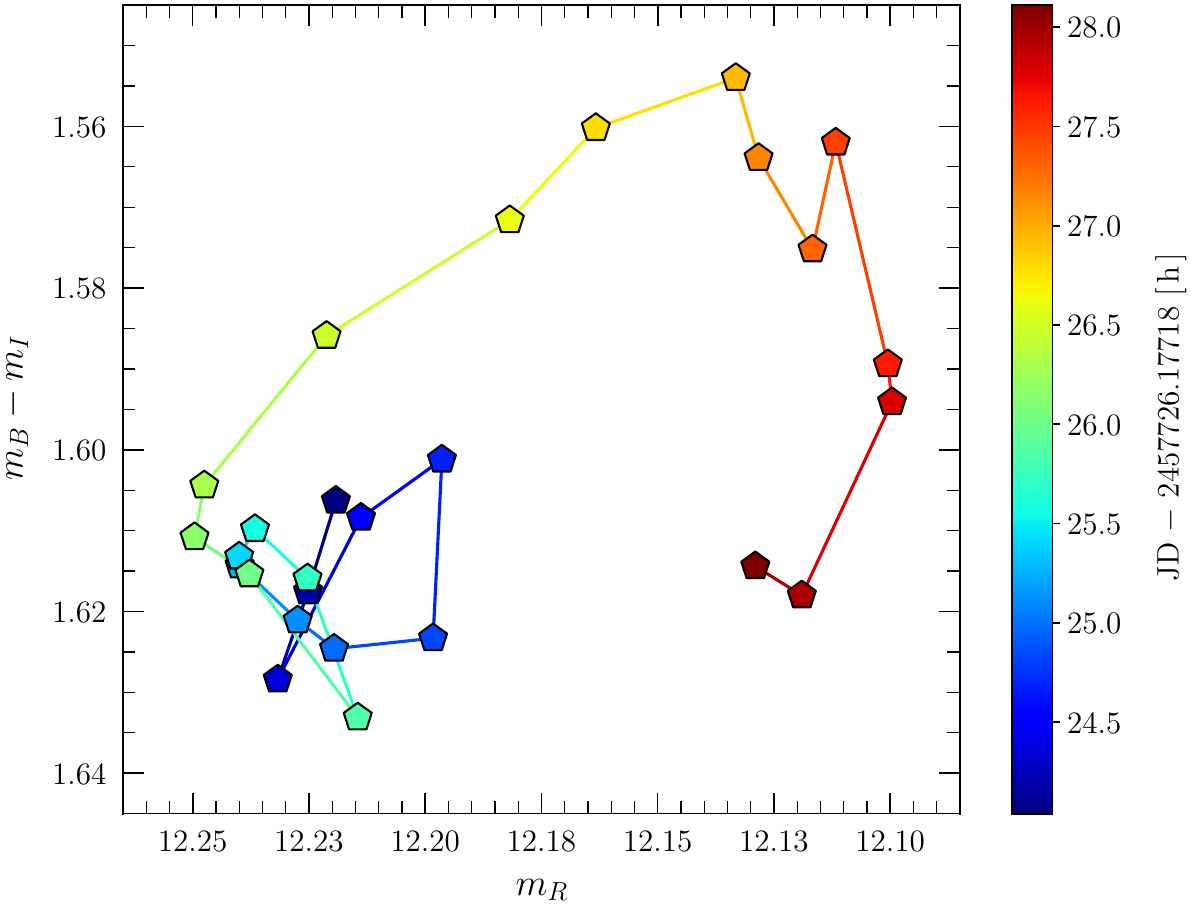}
\vspace{-3mm}
\caption{Colour-magnitude diagram for December 4$^{\rm th}$. A clockwise spectral hysteresis loop can be traced. The different colours denote the different observing times as indicated on the right.}
\label{fig:cmd}
\vspace{-1mm}
\end{figure}

\vskip-3mm

\section{RESULTS AND DISCUSSION}

\noindent The object showed brightness variations in the range 0.1--0.2\,mag over all three nights. During the first night, the variability pattern revealed two components~-- we observed low-amplitude flares, best separated in the $R$ band, superimposed onto a smooth trend with a time scale longer than the monitoring duration. In spite of their low amplitude, we shall consider the flares real since they can be traced in all bands.

\indent The second night started with a weakly decreasing trend with some fluctuations superimposed. Over the second part of the night a flaring event was observed (i.e. there was a two-component variability pattern again). The close inspection of the event revealed a very close superposition of two flares.

\indent During the third night, we observed a flare followed by an almost constant flux level. At the beginning a flux fluctuation was detected, which pointed to preceding variations.

\indent Generally, the intra-night variability pattern reveals two components~-- a smoothly varying component\footnote{Actually, this component represents inter-night (or night-to-night) blazar variability.}, characterized by a time scale longer than the monitoring duration\footnote{We assume that in general, the monitoring lasts over the whole night.} and a flaring component consisting of one or more flaring events, each characterized by a time scale shorter than the monitoring duration. 

\indent In order to get more accurate flare parameters, one needs to remove the smooth component, that is, to do a detrending (see Agarwal et al. 2023 for detailed discussion and examples). For each night, we fitted a first or second degree polynomial to the LC data points, which we assume to belong to the smooth, low-frequency component. For a given night, the degree of polynomial and the fitting interval(s) are one and the same for all bands. The fits are shown in Figure\,\ref{fig:lc}. They are used to detrend the corresponding LCs under the assumption that the smooth component is of geometric origin, that is, we actually did a deboosting. Before that, the dereddened\footnote{Galactic extinctions are taken from NED: $\{A_B,A_V,A_R,A_I\}=\{0.261,0.198,0.156,0.108\}$\,mag.} LCs and polynomials were transformed to fluxes using the zero-points of Bessell et al. (1998). After the deboosting, the flux variations are characterized by a single, but unknown value of the Doppler factor (Figure\,\ref{fig:lc1}).

\indent We built $(m_B-m_I)$ against $m_R$ colour-magnitude diagrams for all nights of monitoring. For the first night, we found no significant chromatism (a linear Pearson correlation coefficient $r = -0.408$ and a probability to get such a coefficient by chance $p = 0.104$). For the third night, we got a significant bluer-when-brighter (BWB) chromatism ($r = 0.747$ and $p = 0.00001$). 

\indent The second night is most intriguing~-- in addition to the significant BWB chromatism ($r = 0.598$ and $p = 0.001$), we detected a clockwise spectral hysteresis loop (the BWB and redder-when-fainter trends form a crude ellipse, Figure\,\ref{fig:cmd}). This is an indication of evolution of the electron energy distribution owing to acceleration and cooling processes.

\indent Let us assume that the flares observed on December 4$^{\rm th}$ are produced by a shock hitting small inhomogeneities/turbulent cells within the jet. This leads to acceleration of electrons at the shock front followed by their cooling in the post-shock region via synchrotron and inverse Compton radiation. The electrons lose half of their energy within the cooling time, $t_{\rm cool}(\nu)$:
\begin{equation}
\centering
t_{\rm cool}(\nu) \simeq \frac{4.73\!\times\!10^4}{1+q}\,({\mathcal B'})^{-3/2}\, \nu_{15}^{-1/2}\, \left(\frac{\delta}{1+z}\right)^{-1/2} \quad [\,\rm s\,],
\end{equation}
\noindent where $\nu_{15}$ is the observed photon frequency (in units of $10^{15}$\,Hz, $\nu = 10^{15}\nu_{15}$\,Hz), ${\mathcal B'}$ is the co-moving magnetic field strength (in units of Gauss), $\delta$ is the Doppler factor, and $q$ is the Compton dominance parameter. The latter is the ratio of the co-moving energy densities of the radiation and magnetic fields, $q=U'_{\rm rad}/U'_{\mathcal B}$.
In the framework of this scenario, the clockwise spectral hysteresis arises when the synchrotron cooling time dominates over the acceleration and light-crossing times (Kirk et al. 1999). In addition, time lags between the flux variations in different bands should be expected~-- the electron Lorentz factor evolves as $\dot{\gamma}_{\rm e} \propto -\gamma^2_{\rm e}$ and, therefore, the most energetic electrons are cooled first followed by the less energetic ones, thereby resulting in the so called soft time lag.

A detailed analysis of the CTA\,102 variability will be presented elsewhere.

\vskip2mm

\centerline{\bf Acknowledgements}

\vskip2mm

\noindent This research was partially supported by the Bulgarian Ministry of Education and Science under Agreement D01-326/04.12.2023.
This research has made use of the NASA/IPAC Extragalactic Database (NED), which is funded by the National Aeronautics and Space Administration and operated by the California Institute of Technology.

\vskip-.5cm

\references

Agarwal, A., Mihov, B., et al.: 2023, \journal{The Astrophysical Journal Suppl. Series}, \vol{265}, id.51.

Bessell, M. S., Castelli, F., Plez, B.: 1998, \journal{Astronomy and Astrophysics}, \vol{333}, 231. 

Kirk, J. G., Rieger, F. M., Mastichiadis, A.: 1999, \journal{ASP Conference Series}, \vol{159}, 325. 

Kostov, A.: 2010, \journal{Proc. of Gaia Follow-up Network for Solar System Objects Workshop} (online at https://gaiafunsso.imcce.fr/workshop/GFSSO-Proceedings-2010.pdf), 137.

Raiteri, C. M., Villata, M., Acosta-Pulido, J. A., et al.: 2017, \journal{Nature}, \vol{552}, 374.

\endreferences

\end{document}